\documentclass[12pt,preprint]{aastex}
\begin{document}

\shortauthors{Mallouris et al.}
\shorttitle{Interstellar Gas toward Sk 108}

\title{{\it FUSE} Observations of Interstellar Gas toward the SMC Star Sk 108}

\author{C. Mallouris\altaffilmark{1}, 
D. E. Welty\altaffilmark{1}, 
D. G. York\altaffilmark{1,2} 
H. W. Moos\altaffilmark{3}, 
K. R. Sembach\altaffilmark{3}, 
S. D. Friedman\altaffilmark{3}, 
E. B. Jenkins\altaffilmark{4}, 
M. Lemoine\altaffilmark{5}, 
W. R. Oegerle\altaffilmark{3}, 
B. D. Savage\altaffilmark{6}, 
J. M. Shull\altaffilmark{7}, 
G. Sonneborn\altaffilmark{8}, and
A. Vidal-Madjar\altaffilmark{9}} 

\altaffiltext{1}{University of Chicago, Department of Astronomy \& Astrophysics, 5640 S. Ellis Ave., Chicago, IL 60637}

\altaffiltext{2}{Enrico Fermi Institute}

\altaffiltext{3}{Department of Physics and Astronomy, Johns Hopkins University 3400 N. Charles St., Baltimore, MD 21218}

\altaffiltext{4}{Princeton University Observatory, Princeton, NJ 08544}

\altaffiltext{5}{D\'{e}partement d'Astrophysique Relativiste et de Cosmologie, UMR-8629 CNRS, Observatoire de Paris-Meudon, Place Jules Janssen, F-92195 Meudon, France}

\altaffiltext{6}{Department of Astronomy, University of Wisconsin, 475 N. Charter St., Madison, WI 53706}

\altaffiltext{7}{Center for Astrophysics and Space Astronomy, Department of Astrophysical and Planetary Sciences, University of Colorado, Campus Box 389, Boulder, CO 80309}

\altaffiltext{8}{Laboratory for Astronomy and Solar Physics, NASA Goddard Space Flight Center, Code 681, Greenbelt, MD 20771}

\altaffiltext{9}{Institut d'Astrophysique de Paris, CNRS, 98 bis Boulevard Arago, F-75014 Paris, France}

\begin{abstract}

We discuss the interstellar absorption lines found in {\em FUSE} spectra of the Wolf-Rayet binary Sk 108, located in the northeastern part of the main ``bar'' of the Small Magellanic Cloud.  
The spectra cover the wavelength range $988-1187$ {\mbox \AA}, at a resolution of about 12,000 and S/N of 20--40.  
We use detailed component information from higher resolution near-UV and optical spectra to model the far-UV lines of similarly distributed species.
Both the Galactic and SMC gas toward Sk 108 seem to be predominantly neutral, though a significant fraction of the SMC gas is ionized.
The column densities of \ion{P}{2}, \ion{S}{2}, and \ion{Ar}{1} are consistent with essentially solar ratios, relative to $N$(\ion{Zn}{2}), in both the Galactic and SMC gas; the column density of \ion{N}{1} remains somewhat uncertain.
Molecular hydrogen is present in the Galactic gas, with properties similar to those found in low mean density Galactic lines of sight and in the Galactic gas toward several other LMC and SMC stars.
We report a tentative detection of H$_2$ in the SMC gas for $J$ = 1 and 3, with rotational level populations consistent with an excitation temperature of order 1000 K --- similar to the H$_2$ found in diffuse Galactic gas toward $\zeta$ Pup.
Strong absorption from \ion{N}{3}, \ion{S}{3}, and \ion{Fe}{3} has revealed a significant ionized component, particularly in the SMC; \ion{O}{6} is present, but relatively weak, especially in the Galactic gas.
The $N$(\ion{C}{4})/$N$(\ion{O}{6}) ratio varies somewhat within the SMC --- suggesting that several processes may contribute to the observed high ion abundances.
\end{abstract}


\section{Introduction}
\label{sec-intro}

The observed sub-solar gas phase abundances [X/H] of most heavy elements in the Galactic interstellar medium (ISM) are usually taken to indicate that the ``missing'' atoms have been depleted into dust grains that generally are coextensive with the gas (e.g., Jenkins 1987; Savage \& Sembach 1996).
Observations of heavy elements in QSO absorption-line systems (ALS), which likely sample the ISM of galaxies of various kinds at various evolutionary stages, may provide the best means of tracing the build-up of those elements for redshifts $z$ $\la$ 5 --- if the possible (or likely!) effects of depletion and ionization can be assessed (e.g., Lu et al. 1996; Lauroesch et al. 1996; Pettini et al. 1997; Prochaska \& Wolfe 1997).
Howk \& Sembach (1999) have explored the effects of the presence of some ionized gas on the elemental abundances inferred for QSOALS.
Several other recent studies have attempted to ascertain the underlying total abundance patterns (and thus the nucleosynthetic histories) of some QSOALS by adjusting the observed gas phase abundances for an assumed level and pattern of depletion (e.g., Kulkarni, Fall, \& Truran 1997; Vladilo 1998).
Such studies depend, however, on the assumption that the dust depletion patterns in the QSOALS resemble those found in our own Galactic ISM --- i.e., that they are independent of metallicity.
Studies of the abundances and physical conditions in the ISM of the Magellanic Clouds --- nearby systems of low metallicity and low dust-to-gas ratio (similar to some QSOALS) for which stellar abundance data are also available --- can provide a significant test of that assumption.
 
UV spectra of the SMC Wolf-Rayet binary Sk 108, obtained with the {\it HST} GHRS, have yielded the most extensive and accurate abundances currently available for the SMC ISM (Welty et al. 1997; hereafter Paper I).
Fits to the GHRS echelle profiles of absorption lines due to \ion{Si}{2}, \ion{Mn}{2}, and \ion{Fe}{2} toward Sk 108 required (at least) 25 components, which were grouped into 5 sets based on kinematics and/or relative abundances. 
One set of components (G1, at $-$6 to 6 km~s$^{-1}$) arises in the Galactic halo, one set (G2, at 13 to 57 km~s$^{-1}$) arises in the Galactic disk, and three sets (S1, at 95 to 156 km~s$^{-1}$; S2, at 167 to 196 km~s$^{-1}$; S3, at 209 to 216 km~s$^{-1}$) are attributable to gas in the SMC. 
The total \ion{H}{1} column densities, for both Galactic and SMC gas, are roughly 3.5~$\times$~10$^{20}$~cm$^{-2}$ (Fitzpatrick 1985), predominantly in the G2 and S1 components. 
The halo components and the three SMC component groups all exhibit relative abundance patterns ([X/Zn], for X = Si, Cr, Mn, Fe, and Ni) similar to those found in Galactic halo clouds; the G2 components have abundances similar to those in warm, diffuse clouds in the Galactic disk.  
The relative gas phase abundances found for four sets of LMC components toward SN 1987A also resemble (on average) those found either in warm Galactic disk clouds or in halo clouds (Welty et al. 1999a; hereafter Paper II).
Since the relative (total) abundances of those elements found in LMC and SMC stars are similar to those in analogous Galactic objects (e.g., Russell \& Dopita 1992; see the appendices to Papers I and II), the depletion patterns may well be very similar in the Galaxy, LMC, and SMC --- at least where the depletions are relatively mild.
Interestingly, the relative gas phase abundance patterns seen for a number of QSOALS also are similar to those found for the Galactic halo clouds and for the SMC ISM toward Sk 108 (Paper I).

Sk 108 was observed with the {\em Far Ultraviolet Spectroscopic Explorer} during the commissioning phase of the {\em FUSE} mission.  
We present here a preliminary analysis of the {\it FUSE} LiF1A and LiF1B spectra of Sk 108, which cover the wavelength range 988--1187 {\mbox \AA} at a resolving power R $\sim$ 12,000.  
These far-UV spectra include absorption lines from a number of species (e.g., H$_2$, \ion{N}{1}, \ion{Ar}{1}, \ion{Fe}{3}, \ion{O}{6}) not observed in the existing GHRS spectra of Sk 108, and therefore provide additional information on the abundances and physical conditions in both neutral and ionized gas along that line of sight.
As the interstellar component structure toward Sk 108 is relatively well understood from the existing optical and near-UV spectra, analysis of the far-UV line profiles also allows us to gauge both the resolution and the wavelength precision characterizing the {\it FUSE} spectra.

In the following sections, we discuss the processing and analysis of the {\it FUSE} spectra (Sec.~\ref{sec-obs}), the various neutral and ionized species observed (Sec.~\ref{sec-species}), and the abundances, depletions, and physical conditions in the Galactic and SMC gas toward Sk 108 (Sec.~\ref{sec-abund}).
We provide a brief summary in Section~\ref{sec-summary}.

\section{Observations and Data Analysis}
\label{sec-obs}

The {\it FUSE} observations of Sk 108 analyzed in this paper were obtained on 1999 October 20. 
Eight spectra, with a total exposure time of 1.33 $\times$ 10$^{4}$ sec, were acquired through the 30\arcsec~ aperture, processed via the standard {\it FUSE} data pipeline (version 1.5.3; Sahnow et al. 2000), aligned, and co-added.
Sections of the individual spectra affected by noise bursts were excised before co-addition.  
No significant differences were found between spectra obtained during orbital night and those obtained during the day, so all the spectra were combined.
As the SiC channels were poorly aligned during these observations, we consider only the LiF channels, which span the wavelength range from 988 to 1187 \AA.
The summed spectra were then re-binned to approximately optimal sampling at the achieved resolution R = $\lambda$/${\Delta\lambda}$ $\sim$ 12,000 ($\Delta v$ $\sim$ 25 km s$^{-1}$; see below), as shown in Figure~\ref{fig:fullspectrum}. 
The LiF 1a segment is roughly centered on the stellar \ion{O}{6} P Cyg profile and the interstellar \ion{H}{1} Lyman $\beta$ absorption line; numerous sharp interstellar and broader stellar absorption lines are evident in both segments; the very broad, shallow depression centered on about 1155 \AA~ is a known optical anomaly (Sahnow et al. 2000).
Regions near absorption lines of interest were normalized via low-order polynomial fits to the adjacent continuum regions.
The S/N in the summed, re-binned spectrum, determined from those continuum fits, ranges from about 20 to 40.  
Examination of saturated lines and comparisons between daytime and night-time data indicate that both scattered light and airglow lines are negligible.  
Additional spectra of Sk 108 obtained subsequent to improvements in focus and channel alignment --- thus characterized by somewhat higher resolving power and including both LiF and SiC channels --- will be discussed in a future paper (Mallouris 2001, in preparation).

The normalized absorption-line profiles were fitted, using descendants of the programs discussed by Vidal-Madjar et al. (1977) and Welty, Hobbs, \& York (1991), in order to obtain column densities for the various species detected in the far-UV spectra.  
In the fits, we assume that the {\it FUSE} instrumental profile may be approximated by a Gaussian function, and we use the rest wavelengths and oscillator strengths tabulated by Morton (1991; with updates in Welty et al. 1999b).
Because of the relatively low spectral resolution achieved, the detailed component structure derived in Paper I from the GHRS echelle spectra (9 Galactic and 16 SMC components, with $b$-values 2.0-4.5 km s$^{-1}$) cannot be discerned in these {\it FUSE} spectra.
We therefore adopted the component structures derived from the higher resolution (FWHM $\sim$ 4.2 km s$^{-1}$) near-UV spectra of \ion{Si}{2}, \ion{Mn}{2}, \ion{Fe}{2}, and \ion{Zn}{2} (representing ranges in depletion behavior) to model the unresolved absorption line profiles of similarly distributed atomic species seen in the {\em FUSE} spectra.
In view of the generally good correlation between $N$(\ion{Na}{1}) and $N$(H$_2$) found in the Galactic ISM (Federman 1981; Welty \& Hobbs 2001), we used the component structure derived from moderately high resolution (R = 60,000) spectra of the \ion{Na}{1} D lines (Welty et al., in preparation) to model the absorption from the numerous transitions of molecular hydrogen, using the H$_2$ rest wavelengths and oscillator strengths listed by Morton \& Dinerstein (1976). 
While we had hoped to at least determine total column densities for each of the five component groups, small-scale uncertainties in the {\it FUSE} wavelength scale (see below) generally limited us to estimates for the total Galactic (G1 + G2) and total SMC (S1 + S2 + S3) column densities.
Consequently, the line widths ($b$-values) of all 25 components and the relative velocities and relative column densities of the Galactic and SMC components (separately) were fixed in the fits --- which thus yielded velocity offsets and total column densities for both the Galactic and the SMC absorption.

The fits to the numerous far-UV \ion{Fe}{2} and H$_2$ profiles also yielded both estimates for the resolution characterizing these {\it FUSE} spectra and corrections to the default wavelength scale.
The detailed component structure for \ion{Fe}{2} should be particularly well determined, as it is based on a mutually consistent fit to GHRS echelle spectra of the lines at 2260 and 2586 \AA, which differ from each other by a factor of 30 in $f\lambda$ (Paper I).
Fits to the profiles of additional \ion{Fe}{2} lines in the far-UV (generally of intermediate strength), assuming that component structure, therefore enabled us to estimate the resolution achieved in the {\it FUSE} spectra:  R $\sim$ 12,000, apparently uniform over the wavelength range of the LiF 1A and 1B channels. 
We note that improvements to the focus in early 2000 have yielded significantly higher spectral resolution.
The velocity offsets (relative to the optical and near-UV spectra) for the H$_2$ and \ion{Fe}{2} lines are shown in Figure~\ref{fig:wave} by crosses and open circles, respectively.\footnotemark
\footnotetext{Comparisons of the new optical profiles of Galactic \ion{Na}{1} with the near-UV profile of the $\lambda$1328 line of \ion{C}{1} suggest that the velocities adopted in Paper I should be increased by 6 km s$^{-1}$ --- which also yields better agreement with the 21 cm spectra of McGee \& Newton (1986).}
The filled circles show the corresponding offsets for an improved wavelength solution implemented in the CALFUSE pipeline in 2000 April.
The error bars represent the (approximately) 1$\sigma$ uncertainties determined in the profile fits.  
For both wavelength solutions, the absolute velocities are systematically too small (common to many {\em FUSE} spectra), and the relative velocities can be uncertain by $\sim$ 10 km s$^{-1}$ over relatively short wavelength intervals (see also Sahnow et al. 2000).
There was no appreciable drift in the velocity offsets between the first and last exposures on Sk 108 --- over a span of eight orbits.
Unless otherwise noted, the {\it FUSE} Early Release Object (ERO) papers (ApJL, vol. 538) use the same (pre-2000 April) scale used here, and so will have the same small scale errors.  
The error in the absolute velocities depends on the centering of the target in the 30\arcsec~ aperture used for all ERO observations, and thus may differ for different data sets.  
Further improvements to the {\it FUSE} wavelength scale are being pursued.

\section{Observed Species}
\label{sec-species}

\subsection{Neutral and Singly Ionized Atomic Species}
\label{sec-atomic}

The {\it FUSE} spectra of Sk 108 exhibit absorption lines from a number of species which are dominant ions in primarily neutral (\ion{H}{1}) gas.  
The adopted atomic data, the measured equivalent widths, and the column densities derived from fits to the individual line profiles for some of these are listed in Table~\ref{tab:neutral}. 
Some of the corresponding absorption-line profiles are shown in Figure~\ref{fig:neutral}, where the absolute velocities for the {\it FUSE} spectra have been determined in the profile fits. 
The reference abundances, the adopted column densities $N$(X) and the relative abundances [X/Zn], for both the Galactic and SMC gas, are listed in Table~\ref{tab:abund}. 
Because N, P, and Ar are generally at most mildly depleted in the Galactic ISM (Meyer, Cardelli, \& Sofia 1997; Jenkins 1987; Sofia \& Jenkins 1998), we used the \ion{Zn}{2} component structure (Paper I) in fitting the lines for \ion{N}{1}, \ion{Ar}{1} and \ion{P}{2}.
As the \ion{Zn}{2} $\lambda$2062 line is weaker than those far-UV lines, however, we have augmented the \ion{Zn}{2} component structure to include additional components seen in the strong \ion{Fe}{2} $\lambda$2586 line, with the \ion{Zn}{2} column densities estimated via typical values of $N$(\ion{Zn}{2})/$N$(\ion{Fe}{2}) for each of the five component groups.
While the low spectral resolution and the small-scale uncertainties in the wavelength scale generally make it difficult to determine accurate abundances for each of the five component groups, the line profiles do in some cases suggest differences in abundance ratios among those groups.

\ion{Fe}{2}:  
Nine \ion{Fe}{2} lines were observed, with 0.45 $\lesssim$ log($f\lambda$) $\lesssim$ 2.08 and at wavelengths from 1055 to 1144 {\mbox \AA}.  
We fitted the line profiles using the component structures determined from the GHRS echelle spectra of the $\lambda$2260 and $\lambda$2586 lines of \ion{Fe}{2}, with log($f\lambda$) = 0.74 and 2.22, respectively (Paper I).  
As the \ion{Fe}{2} column densities should be well determined from the GHRS spectra, we used the fits to the far-UV \ion{Fe}{2} lines to check the corresponding $f$-values (which in many cases are less well known).
Howk et al. (2000) have combined \ion{Fe}{2} spectra from {\it Copernicus}, GHRS, and {\it FUSE} to determine empirical $f$-values for eleven far-UV \ion{Fe}{2} transitions [see also Shull, Van Steenberg, \& Seab (1983); de Boer et al. (1974)].
The $f$-values inferred from our fits to the {\it FUSE} spectra of Sk 108 (last column of Table~\ref{tab:neutral}) agree with those derived by Howk et al. for the lines at 1055, 1121, 1125, 1142, and 1143 \AA, but are lower by 30--40\% for the lines at 1096, 1112, 1133, and 1144 \AA~ [though all but $f$(1096) and $f$(1144) may be consistent within the mutual uncertainties].
On average, the $f$-values estimated here are about 20\% lower than those of Howk et al. (but are generally within the range of values found for individual lines of sight); the differences do not appear to depend in any systematic way on line strength.

\ion{P}{2}:  
The $\lambda$1152 line of \ion{P}{2} provides a measure of the phosphorus abundance in the SMC ISM,
as \ion{P}{2} $\lambda$1301 is blended with the much stronger \ion{O}{1} $\lambda$1302 line in the GHRS G160M spectrum (Paper I).  
We used the augmented \ion{Zn}{2} component structure to fit the \ion{P}{2} lines;
uniform $N$(\ion{P}{2})/$N$(\ion{Zn}{2}) ratios yield quite acceptable fits to both the Galactic and SMC \ion{P}{2} absorption.
The Galactic \ion{P}{2} column density toward Sk 108 derived from the $\lambda$1152 line is a factor of 2 lower than that found from the $\lambda$1301 line, however.
While this difference might be suggestive of saturation and/or component structure effects (see below) in the stronger $\lambda$1152 line, we note that a similar difference was found for the line of sight toward 23 Ori (Welty et al. 1999b), that such differences are within the uncertainties in the $f$-values (Morton 1991)\footnotemark, and that in any case the weaker SMC absorption will be less saturated.
\footnotetext{Morton (2000) lists $f$(1301) = 0.0127 --- smaller than the value 0.0173 in his earlier compilation.  Use of the new $f$-value would exacerbate the differences in column densities inferred from the $\lambda$1152 and $\lambda$1301 lines; the stronger $\lambda$1152 line may thus be somewhat saturated.}
If we increase the column densities derived from the $\lambda$1152 line by 0.15 dex (taking average values from the two lines), the resulting relative abundances [P/Zn] are essentially solar in both the Galactic and SMC gas.  

\ion{Si}{2}:  
Two lines of \ion{Si}{2} ($\lambda$989 and $\lambda$1020) are present in the {\em FUSE} spectrum.  
Since both are considerably stronger than the $\lambda$1808 line measured with GHRS, we have estimated $N$(\ion{Si}{2}) for the components seen only in the stronger \ion{Fe}{2} $\lambda$2586 line via the column density ratios measured for the other components (Paper I).  
Even with that augmentation of the \ion{Si}{2} structure, however, the column densities derived from fits to the $\lambda$1020 line are factors of 2.5--3.5 smaller, for the SMC and Galactic components, than those found from fitting the weaker $\lambda$1808 line.
While some of the difference may be ascribed to uncertainties in the $f$-value for the $\lambda$1020 line\footnotemark, saturation effects may also be present. 
\footnotetext{Morton (2000) lists $f$(1020) = 0.0164 --- smaller than the value 0.0282 in his earlier compilation --- which would reduce the differences between the column densities inferred from the $\lambda$1020 and $\lambda$1808 lines to factors of 1.5--2.0.}  
The $\lambda$989 line of \ion{Si}{2} is severely blended with \ion{N}{3}.

\ion{Ar}{1}:  
We used the augmented \ion{Zn}{2} component structure to fit the \ion{Ar}{1} lines at 1048 and 1066 \AA. 
The stronger $\lambda$1048 line is essentially free of contamination by other features, but the weaker $\lambda$1066 line is blended with two H$_{2}$ lines and with a broad stellar line of \ion{Si}{4} (Morton \& Underhill 1977).  
Fits to the weaker line yield \ion{Ar}{1} column densities about 0.1 dex higher than the fits to the stronger line, suggestive of mild saturation effects.
While a uniform $N$(\ion{Ar}{1})/$N$(\ion{Zn}{2}) ratio yields good fits to the Galactic \ion{Ar}{1} absorption,
some of the weaker outlying SMC components (10--12, 21--25) may have somewhat smaller such ratios than the rest of the SMC components.  
The SMC column densities in Table~\ref{tab:neutral} were derived assuming $N$(\ion{Ar}{1})/$N$(\ion{Zn}{2}) is lower by a factor 10 in those lower column density   components, where Ar may be partially ionized (see Sec.~\ref{sec-neutral}).
If the $b$-values for the individual components were smaller, the derived column densities would be somewhat higher.

\ion{N}{1}:  
The three lines in the \ion{N}{1} triplet at 1134 {\mbox \AA} are strong and severely blended with each other -- making it difficult to derive reliable column densities.  
The initial fits, using the augmented \ion{Zn}{2} component structure, seemed to suggest that [N/Zn] is significantly less than solar for both the Galactic and SMC gas ($-0.8$ to $-1.05$ dex). 
The lines are very saturated, however, so the derived column densities should be considered as lower limits.  
In addition, the empirical curves of growth for \ion{N}{1} (both Galactic and SMC) appear to have smaller effective $b$-values than the curves followed by most singly ionized species, and the profiles of the (unblended) Galactic \ion{N}{1} $\lambda$1134.1 and SMC \ion{N}{1} $\lambda$1134.9 lines appear to be slightly narrower than the profiles of comparably strong lines from \ion{O}{1} and various singly ionized species (Fig.~\ref{fig:neutral}; see also Sec.~\ref{sec-neutral}).
We therefore also tried fits in which the outlying Galactic and SMC components (1, 7--9, 10--12, 21--25) have smaller $N$(\ion{N}{1})/$N$(\ion{Zn}{2}) ratios by a factor 10.
Such fits yielded both better agreement with the observed line profiles and somewhat larger overall \ion{N}{1} column densities.
Observations of some of the weaker \ion{N}{1} multiplets below 980 \AA~ will enable tighter limits to be placed on $N$(\ion{N}{1}).

For each of the species \ion{N}{1}, \ion{Si}{2}, \ion{P}{2}, and \ion{Ar}{1}, there appears to be a systematic trend for lower column densities to be derived from the stronger far-UV lines, especially for the Galactic absorption.
While that is not the case for \ion{Fe}{2} (the adopted component structure yields good fits to both the weak $\lambda$2260 and the strong $\lambda$2586 lines), lines from less severely depleted species may follow a somewhat different curve of growth (with different relative abundances in the various components) which saturates at lower overall column densities (cf. Welty et al. 1999b).
As an additional test, we fitted the GHRS G160M profiles (R $\sim$ 14,000) of the three \ion{S}{2} lines at 1250, 1253, and 1259 \AA~ (which range over a factor of about 3 in strength), using the augmented \ion{Zn}{2} component structure.
Again, the weakest line yielded the highest column densities, for both Galactic and SMC components (Table~\ref{tab:neutral}), and even the values derived from the $\lambda$1250 line imply slightly lower [S/Zn] ratios than would be expected if the depletions are similar to those found in warm disk clouds (+0.2 dex) or halo clouds (+0.1 dex) (Papers I and II).
We therefore suspect that the column densities derived from fits to the stronger UV lines of these mildly depleted species may be somewhat underestimated.
As the $b$-values found in the fits to the GHRS echelle spectra are relatively large ($>$ 3 km s$^{-1}$), the true structure may be more complex --- more components, with smaller $b$-values.
Alternatively, the fits to the \ion{Ar}{1} and \ion{N}{1} lines suggest that fewer components (perhaps with smaller $b$-values) may contribute to those neutral species.
In either case, the true column densities could well be higher.
Higher resolution spectra of lines from \ion{N}{1}, \ion{O}{1}, and \ion{S}{2} would help to define the appropriate component structure for the stronger lines of species with very mild depletions.
Finally, we note that single component fits to the Galactic absorption (with $b$ $\sim$ 10--15 km s$^{-1}$) yield column densities up to a factor 2 smaller than those found using the detailed component structure; the corresponding discrepancies for single component fits to the weaker, less saturated SMC absorption (with $b$ $\sim$ 20--27 km s$^{-1}$) are somewhat smaller.

\subsection{Molecular Hydrogen}
\label{sec-h2}

Many absorption features due to Galactic molecular hydrogen are apparent in the {\em FUSE} spectrum of Sk 108.  
We have analyzed lines from the Lyman bands from 0-0 to 9-0; lines from the Werner 0-0 band are also present.
In Figure~\ref{fig:h2}, we show the profiles for rotational levels $J$ = 0, 1, 2, and 3 --- in each case, averaged over three lines of similar strength (as noted in the caption).
At the top is the higher resolution \ion{Na}{1} spectrum, showing more of the details of the component structure.  
As the $b$-values determined for the five \ion{Na}{1} components used to model the various Galactic H$_{2}$ lines are typically about 1.0 km s$^{-1}$, the gas must generally be cooler than about 1500 K.  
We have therefore fitted those H$_{2}$ lines assuming $T$ = 100 K (typical for cool neutral disk gas) and also assuming $T$ = 1200 K.
In each case, the non-thermal contribution to $b$(H$_2$) was assumed to be the same as for \ion{Na}{1}.
The derived column densities, listed in Table~\ref{tab:h2}, are higher for $T$ = 100 K (the more likely value, given the rotational temperature noted below).  
The lines from $J$ = 0 and 1 are relatively strong, and lie essentially on the flat part of the curve of growth --- so that the column densities are not well determined (as reflected in the differences between the values derived for different $T$).  
At least some of the lines from the $J$ = 2 and 3 levels are weak enough, however, to permit reliable column density determination.
Comparison of the H$_2$ rotational populations in the Galactic gas toward Sk 108 with those seen in several low mean density Galactic lines of sight ($\mu$ Col, HD 93521; Table~\ref{tab:h2}) shows that $N$($J$) follows a common pattern, in which $N$(1) $>$ $N$(0) $>$ $N$(2) $>$ $N$(3) $>$ $N$(4).
The rotational temperature $T_{01}$ is of order 100 K, while the ``excitation temperatures'' $T_{\rm ex}$ obtained by comparing the column densities for $J$ = 3 vs. $J$ = 1 and for $J$ = 4 vs. $J$ = 2 are of order 200--300 K.  
The total $N$(H$_2$) is consistent with the general Galactic relationships between $N$(H$_2$), $N$(\ion{Na}{1}), and $N$(H) (Welty \& Hobbs 2001), and also with the values observed for the Galactic gas toward two other SMC stars (Shull et al. 2000).

Absorption from H$_2$ is evidently much weaker in the SMC gas toward Sk 108 --- even though the \ion{H}{1} column density is comparable to that in the Galactic gas.
We do not detect any H$_{2}$ absorption at SMC velocities for $J$ = 0 or 2.
Several of the lines from $J$ = 3 and $J$ = 1, however, appear to exhibit weak, broad absorption centered at $v$ $\sim$ 137 km s$^{-1}$ --- similar to the mean velocity of the S1 components.
Simultaneous one component fits to the available unblended H$_2$ lines yield very similar velocities and widths for both $J$ = 1 and $J$ = 3.
The column densities for $J$ = 1 and $J$ = 3 are both of order 10$^{14}$ cm$^{-2}$; the 3$\sigma$ limits for $J$ = 0, $J$ = 2, $J$ = 4, and $J$ = 5 are all about 0.4--0.6 $\times$ 10$^{14}$ cm$^{-2}$.  
Comparison with the Galactic lines of sight listed in Table~\ref{tab:h2} shows several similar cases (HD 28947 A and $\zeta$ Pup) where the pattern of column densities is $N$(3) $>$ $N$(1) $>$ $N$(2) $>$ $N$(4) $>$ $N$(0). 
In such cases, all the rotational levels can be characterized by a single relatively high excitation temperature of order 1000 K; the lines from $J$ = 3 are strongest due to the high statistical weight.
If $T_{\rm ex}$ = 1000 K, then the total $N$(H$_2$) in the SMC gas toward Sk 108 is roughly 3 $\times$ 10$^{14}$ cm$^{-2}$ --- somewhat smaller than the values found for the SMC gas toward HD 5980 and AV 232, which have slightly higher $N$(H), somewhat lower $T_{\rm ex}$, and $T_{01}$ $<$ $T_{\rm ex}$ (Shull et al. 2000).  
In a subsequent survey of H$_2$ in the Magellanic Clouds (Tumlinson et al. 2000), H$_{2}$ is detected in 24 of the 26 SMC sightlines; the mean fraction of hydrogen in molecular form $<$$f$(H$_2$)$>$ is smaller in the SMC, relative to comparable Milky Way sightlines, by a factor of about five.

\subsection{More Highly Ionized Species}
\label{sec-ionized}

While most of the gas toward Sk 108 appears to be largely neutral, some ionized gas is also present.  
Relatively weak, broad absorption from \ion{Al}{3} and much stronger absorption from \ion{Si}{4} and \ion{C}{4} were detected in spectra from {\em IUE} (de Boer \& Savage 1980; Fitzpatrick \& Savage 1985; Paper I) and GHRS (Hutchings, Bianchi, \& Morris 1993).  
The {\em FUSE} spectra of Sk 108 reveal strong absorption --- especially at SMC velocities --- from \ion{N}{3} ($\lambda$989), \ion{S}{3} ($\lambda$1012), and \ion{Fe}{3} ($\lambda$1122), and somewhat weaker absorption from \ion{O}{6} ($\lambda$1031 and $\lambda$1037).  
Most of these lines are shown in Figure~\ref{fig:ionized}; equivalent widths are given in Table~\ref{tab:ionized}.  

Stellar photospheric and/or wind absorption features, due to a number of high ions, are also present in the {\it FUSE} spectra of Sk 108 --- and can be blended with interstellar absorption from the same or other species.
The photospheric lines --- from \ion{C}{3}, \ion{C}{4}, \ion{N}{3}, \ion{N}{4}, \ion{Si}{4}, \ion{P}{5}, \ion{S}{3}, \ion{S}{4}, and perhaps \ion{O}{3}, \ion{Si}{3}, and \ion{P}{4} (Morton \& Underhill 1977) --- are generally relatively weak, with FWHM of order 200 km s$^{-1}$, and are centered near the systemic velocity of 170 km s$^{-1}$.
The wind lines, seen in \ion{C}{4}, \ion{N}{5}, and \ion{O}{6}, are much broader (the \ion{C}{4} terminal velocity is about 2000 km s$^{-1}$), and exhibit orbital phase dependent variations in strength (Hutchings et al. 1993).
While relatively weak photospheric absorption features were easily removed from the spectra of \ion{Si}{4}, \ion{S}{3}, and \ion{Fe}{3} in Figure~\ref{fig:ionized}, strong stellar absorption features (and several strong interstellar lines) near the \ion{O}{6} doublet at 1031 and 1037 \AA~ have made it very difficult to determine the true amount of interstellar \ion{O}{6} absorption toward Sk 108.
Figure~\ref{fig:o6} shows the region around the two \ion{O}{6} lines, with several possible choices for the shape of the ``continuum'' seen by the interstellar lines.
The corresponding normalized profiles are shown in Figure~\ref{fig:ionized}; the ranges in equivalent width and column density for the \ion{O}{6} lines are given in Table~\ref{tab:ionized}.
The blue-ward extent of the Galactic \ion{O}{6} absorption is primarily constrained for the $\lambda$1031 line, by requiring that the SMC H$_2$ 6-0 P(3) absorption be consistent with that from other lines from $J$ = 3.
The red-ward extent of the SMC \ion{O}{6} absorption is primarily constrained for the $\lambda$1037 line, by requiring similar consistency for the Galactic H$_2$ 5-0 R(2) line.  
As narrow features in the stellar wind lines might, in principle, overlap the Galactic or SMC interstellar \ion{O}{6} lines, the equivalent widths and column densities listed in Table~\ref{tab:ionized} should perhaps be viewed as upper limits.
More definitive interstellar \ion{O}{6} abundances will require a mutually consistent fit to the two profiles, including the overall P Cyg profile, the narrower wind/photospheric features, and the other interstellar lines (H$_2$, \ion{C}{2}, \ion{C}{2}*, \ion{O}{1}).

While the interstellar absorption from the higher ions generally covers a similar velocity range to that of the neutral and singly ionized species, the centroid of the SMC high ion absorption appears to be at somewhat higher velocities (especially for \ion{O}{6}).
We do not know the appropriate component structure(s) for these higher ions, but lower limits to the total Galactic and SMC column densities may be obtained via ``apparent optical depth'' (AOD) integrals (Table~\ref{tab:ionized}).  
If the line widths for these ions are large enough (and if the lines are not too saturated), then those limits may be reasonably close to the actual values.  
Trial fits to the high ion line profiles suggest that at least several SMC components are likely to be present, for ``reasonable'' $b$-values; the total column densities obtained from those fits are typically slightly higher than the AOD limits.

\section{Abundances, Depletions, and Physical Conditions}
\label{sec-abund}

\subsection{Predominantly Neutral Gas}
\label{sec-neutral}

The ions \ion{Zn}{2}, \ion{P}{2}, and \ion{S}{2} are the dominant ionization states of Zn, P, and S in both neutral and ionized gas (as long as the ionization is not too high). 
The abundances of those ions observed toward Sk 108, relative to \ion{H}{1}, suggest, however, that the gas is predominantly neutral. 
For the Galactic components, the abundances are consistent with a small amount of depletion (less than a factor of 2); the depletions would be more severe than those observed in cold, dense clouds if an appreciable amount of ionized (\ion{H}{2}) gas were present.
The same argument holds for the SMC components, when allowance is made for the lower metallicity in the SMC (0.6--0.7 dex below solar).
As discussed in Paper I, the abundances of \ion{Cr}{2}, \ion{Mn}{2}, \ion{Fe}{2}, and \ion{Ni}{2} then indicate that the gas is characterized by ``warm, diffuse cloud'' depletions (G2) or ``halo cloud'' depletions (G1, S1, S2, S3).
There are, however, indications that some ionized gas is also present, particularly in the SMC.
Weak, broad absorption from \ion{Al}{3} and \ion{Si}{2}* was noted in Paper I, and comparisons of $N$(\ion{O}{1}) with the column densities of various singly ionized species suggested that the S3 components (with total $N$(H) of order 10$^{19}$ cm$^{-2}$) might be roughly half ionized.

In principle, the {\it FUSE} data for \ion{Ar}{1}, \ion{N}{1}, \ion{N}{3}, \ion{S}{3}, and \ion{Fe}{3} can provide additional information on the ionization of the Galactic and SMC gas.
Both \ion{N}{1} and \ion{Ar}{1} are dominant ions in neutral gas, but are trace ions in ionized regions; neither N nor Ar is expected to be significantly depleted into dust grains (Meyer et al. 1997; Sofia \& Jenkins 1998).
\ion{N}{1} is a fairly good tracer of \ion{H}{1}, as the ionization of N and H are generally closely linked via charge exchange --- except in moderately ionized gas at very low densities (Jenkins et al. 2000).
Because of its large photoionization cross section in the far-UV, \ion{Ar}{1} can be ionized (and thus deficient) in low column density clouds [$N$(H) $\la$ 10$^{19}$ cm$^{-2}$] that can be penetrated by EUV ionizing radiation (Sofia \& Jenkins 1998; Jenkins et al. 2000).
The individual components detected in \ion{Zn}{2} (Galactic and SMC) are all likely to have $N$(H) $\ga$ 10$^{19}$ cm$^{-2}$, so that Ar and N should be largely neutral there.
Argon (and perhaps nitrogen) could be somewhat ionized, however, in the outlying lower column density components --- particularly in the SMC, where the interstellar radiation field is generally stronger than the average Galactic field (Lequeux 1989).
If there is a substantial amount of ionized gas, we would expect to see deficiences of \ion{N}{1} and \ion{Ar}{1} relative to \ion{Zn}{2} and \ion{S}{2} (and other singly ionized species), but {\it not} relative to \ion{H}{1}.
If the true abundances of \ion{N}{1} (Galactic and SMC) and \ion{Ar}{1} (Galactic) are only slightly above the lower limits listed in Table~\ref{tab:abund}, however, then they would be deficient relative to {\it both} \ion{Zn}{2} and \ion{H}{1}.
In the Galactic gas, the column densities of the trace neutral species \ion{Na}{1} and \ion{C}{1}, relative to $N$(\ion{H}{1}) --- which should be more sensitive to ionization effects --- are quite consistent with the typical relationships seen for predominantly neutral gas (Welty \& Hobbs 2001).
We would argue, therefore, that the Galactic \ion{N}{1} and \ion{Ar}{1} abundances are more likely to be well above the listed limits (so that N and Ar, as well as P and S, have essentially solar abundances, relative to Zn), and that at least the main Galactic components are predominantly neutral.

In the SMC, the overall $N$(\ion{Ar}{1})/$N$(\ion{Zn}{2}) ratio is consistent with the relative reference abundances of Ar and Zn in the SMC (and with solar relative abundances) --- suggestive of (at most) mild depletion for both elements in primarily neutral gas.
As noted above, however, somewhat smaller $N$(\ion{Ar}{1})/$N$(\ion{Zn}{2}) and $N$(\ion{N}{1})/$N$(\ion{Zn}{2}) ratios in some of the weaker outlying SMC components suggest that Ar and N may be slightly ionized there.
While the low overall $N$(\ion{N}{1})/$N$(\ion{Zn}{2}) ratio might seem to imply that N is underabundant in the SMC ISM, the $\lambda$1134 lines are saturated [so $N$(\ion{N}{1}) could be higher].
Furthermore, we note that the reference abundance of N in the SMC is uncertain.
Using values measured for \ion{H}{2} regions and supernova remnants, Russell \& Dopita (1992) adopted A(N)$_{\rm SMC}$ = 6.63 dex --- a factor of about 20 smaller than the solar nitrogen abundance --- whereas most other elements in the SMC are underabundant by factors of 4--5 relative to solar.
More recent SMC stellar abundance studies seem to indicate somewhat higher N abundances --- of order 7.3  dex --- but those stellar values may reflect enriched N (Paper I, and references therein; Venn 1999; Korn et al. 2000).
Another way to constrain the interstellar N abundance is to compare the \ion{N}{1} absorption-line profiles with those of similar strength due to other elements whose abundances are more well determined.
For example, if N/O is solar, then the \ion{O}{1} $\lambda$1039 line will be a factor of about 1.5 stronger (in $N f \lambda$) than the \ion{N}{1} $\lambda$1134.9 line (assuming also that the two species are similarly distributed).
In Figure~\ref{fig:neutral}, however, the SMC \ion{O}{1} $\lambda$1039 line appears to be much stronger than the corresponding \ion{N}{1} $\lambda$1134.9 line --- suggesting either that N is underabundant, relative to O, in the SMC ISM, and/or that N is preferentially ionized, relative to O, in the weaker outlying components.
We note that a tentative detection of the weak \ion{O}{1} $\lambda$1355 line in the SMC gas toward HD 5980 (Koenigsberger et al. 2000) would imply [O/H] $\sim$ $-$0.9 dex --- roughly consistent with the overall SMC metallicity and a ``typical'' oxygen depletion.
The recently obtained {\it FUSE} spectra of Sk 108, which include other \ion{N}{1} lines at shorter wavelengths, should enable more accurate $N$(\ion{N}{1}) to be determined.

For the Galactic gas, the relatively mild depletions, the low $f$(H$_2$), the H$_{2}$ rotational temperature $T_{01}$ $\sim$ 100 K, the slightly higher excitation temperatures for the higher H$_2$ rotational levels (of order 200$-$300 K), and the non-detection of excited state lines of \ion{C}{1} (Paper I) all seem consistent with cool, predominantly neutral, low density gas bathed in a somewhat lower than average ambient radiation field (see, e.g., the H$_2$ calculations of Spitzer \& Zweibel 1974).  
The still milder depletions, lower $f$(H$_2$), higher H$_2$ excitation temperature, and non-detection of \ion{Na}{1} and \ion{C}{1} suggest that the SMC gas may be somewhat warmer, predominantly neutral, low density gas in a somewhat stronger radiation field.
For example, $N$(\ion{Na}{1}) is at least a factor 40 higher in the Galactic gas than in the SMC gas (Welty et al., in preparation) --- even though the hydrogen column densities are comparable.
The difference in overall Na abundance is only a factor of 4--5, so that $\Gamma$/($n_e$ $\alpha$) must be $\ga$ 8 times larger in the SMC gas, where $\Gamma$ is the photoionization rate and $\alpha$ is the recombination rate, roughly proportional to $T^{-0.7}$.

In the Galactic ISM, there appears to be a relationship between the fraction of hydrogen in molecular form $f$(H$_2$) and the degree of the elemental depletions, such that high molecular fractions are associated with more severe depletions (e.g., Cardelli 1994).
Although that relationship is usually taken to indicate a relationship between the depletions and the local hydrogen densities [assuming that high $f$(H$_2$) signifies high $n_{\rm H}$], $f$(H$_2$) also depends on the ambient radiation field, shielding in the lines, and (according to most models of H$_2$ formation) on the grain surface area available for formation of H$_2$. 
For sightlines in the Sco-Oph region, for example, where the radiation field is likely stronger than the typical Galactic field, the $f$(H$_2$)-depletion relationship is offset toward smaller $f$(H$_2$), presumably due to increased photodissociation of the H$_2$ (Welty \& Hobbs 2001).
On the other hand, the shallow far-UV extinction observed for some sightlines in that region suggests that the dust grains may be typically larger there (perhaps due to coagulation of the smaller grains) --- and thus offer a smaller total surface area for H$_2$ formation.
For elements which can be severely depleted (e.g., Ti, Fe), the depletions can range over a factor of 10 or more at any given $f$(H$_2$) --- suggesting that other factors are at work.
Extending such studies to the LMC and SMC (with lower metallicities, lower dust-to-gas ratios, steeper far-UV extinction, and typically stronger radiation fields) may aid in elucidating the relationship between $f$(H$_2$) and depletions --- i.e., is that relationship due primarily to the formation of H$_2$ on grain surfaces, to radiation field effects, or to density-dependent depletion?
The few data for depletions and $f$(H$_2$) available for the LMC and SMC fall within the range observed for Galactic lines of sight, but with generally mild depletions (Papers I and II; Friedman et al. 2000; Shull et al. 2000).
In the SMC, both the lower average $f$(H$_2$) and the shift to higher $N$(H$_{\rm tot}$) for the transition from low to high $f$(H$_2$) (Tumlinson et al. 2000) seem consistent with a generally stronger radiation field.
UV spectra of additional LMC and SMC lines of sight --- particularly with higher $N$(H) and $E(B-V)$ --- would be valuable.

\subsection{Ionized Gas}
\label{sec-ions3}

The detection of absorption from several doubly ionized species provides new information on the ionized gas toward Sk 108.
In Paper I, we estimated $N$(\ion{Al}{3})/$N$(\ion{Al}{2}) to be roughly 0.1 and 0.2 in the Galactic and SMC components, respectively, toward Sk 108.
The column densities for \ion{Fe}{2} and \ion{Fe}{3} in Tables~\ref{tab:neutral}~and~\ref{tab:ionized} suggest that $N$(\ion{Fe}{3})/$N$(\ion{Fe}{2}) is about 0.1 in the Galactic gas, but is $\ga$ 0.5 in the SMC.
Similarly, $N$(\ion{S}{3})/$N$(\ion{S}{2}) is less than about 0.1 in the Galactic gas, but is of order 0.3 in the SMC.
All three ratios for the Galactic gas suggest a fairly low fraction of ionized gas, as argued above.
The larger ratios for the SMC gas are qualitatively consistent with the models of partially (photo)ionized gas discussed by Sembach et al. (1999), in that the ratio for Fe is the largest of the three. 
For $N$(\ion{Fe}{3})/$N$(\ion{Fe}{2}) $\sim$ 0.5, the gas would be roughly half ionized, but the ratios for Al and S are somewhat larger than the models would predict.
Comparison of the profiles of the doubly and singly ionized species in the SMC gas suggests that the $N$(\ion{X}{3})/$N$(\ion{X}{2}) ratios are somewhat higher at velocities corresponding to the S2 and S3 components than at those corresponding to the S1 components.
Given the relatively low resolution of the spectra, however, it is difficult to determine what fractions of the various singly ionized species are associated with primarily neutral gas and what fractions are associated with more ionized gas.

\subsection{High Ions}
\label{sec-high}

Ten of the eleven lines of sight probing the Galactic halo reported by Savage et al. (2000) have equivalent widths for the \ion{O}{6} $\lambda$1031 line (for relatively low-velocity gas in the disk and halo) greater than 100 m\AA, with corresponding \ion{O}{6} column densities greater than 1.4 $\times$ 10$^{14}$ cm$^{-2}$; typical values are 2--3 times higher.
The Galactic \ion{O}{6} $\lambda$1031 line toward Sk 108, however, may be up to a factor 2 weaker than that lower limit --- and at least a factor 2 weaker than the absorption seen toward Sk 80 (AV 232), less than 20 arcmin away (Howk, priv. comm.) --- underscoring the irregularity in the distribution of \ion{O}{6} noted by Savage et al.
The Galactic \ion{O}{6} equivalent width and column density toward Sk 108 are more similar to the values found toward lightly reddened lower Galactic halo stars at 0.5 kpc $\la$ z $\la$ 1.5 kpc, such as HD 28497, $\mu$ Col, and HD 93521 (Shull \& York 1977; York 1977; Caldwell 1979).
The corresponding Galactic absorption from \ion{C}{4}, however, is very strong --- roughly twice as strong as that seen toward Sk 78 and Sk 80, as determined from AOD integrals over the available {\it IUE} spectra.
The $N$(\ion{C}{4})/$N$(\ion{O}{6}) ratio in the Galactic gas toward Sk 108 therefore appears to be significantly larger (3--6) than values typically found for Galactic halo (0.35--1.7) and Galactic disk (0.06--0.28) lines of sight (Spitzer 1996; Savage et al. 2000)\footnotemark.
\footnotetext{We note that the Galactic \ion{O}{6} $\lambda$1031 absorption appears to be roughly twice as strong in the more recent {\it FUSE} spectrum of Sk 108; the reason for this difference is not understood.}  

The strength of the SMC \ion{O}{6} $\lambda$1031 absorption is somewhat uncertain, due to its location near the steeply rising red-ward wing of the stellar wind line, but appears to be comparable to many of the Galactic features in the Savage et al. mini-survey.  
The corresponding SMC \ion{C}{4} absorption, similar in strength to that seen toward Sk 78 and Sk 80, yields an overall SMC $N$(\ion{C}{4})/$N$(\ion{O}{6}) ratio of 1.0--1.6 toward Sk 108 --- within the upper part of the range seen in Galactic halo gas.
While the low resolution and residual wavelength uncertainties make detailed intercomparison of the profiles difficult, the SMC \ion{O}{6} line is clearly shifted red-ward of the centroid of \ion{C}{4}.
Furthermore, the widths of the \ion{O}{6} and \ion{C}{4} absorption (FWHM of order 80 km s$^{-1}$) are much larger than the expected thermal widths --- suggesting that at least several highly ionized components must be present in the SMC gas.
The $N$(\ion{C}{4})/$N$(\ion{O}{6}) ratio thus may vary significantly within the SMC, so that
several of the sources of hot, ionized gas discussed by Spitzer (conductive heating, radiative cooling, turbulent mixing) may be relevant.

\section{Summary}
\label{sec-summary}

We have presented a preliminary analysis of interstellar absorption lines in {\em FUSE} commissioning phase spectra of the SMC Wolf-Rayet star Sk 108.  
The spectra cover the wavelength range 988--1187 \AA, at a resolution of about 12,000 and S/N ranging from 20--40.
We used detailed component information previously derived from GHRS echelle spectra of \ion{Si}{2}, \ion{Mn}{2}, \ion{Fe}{2}, and \ion{Zn}{2} to model the lower resolution {\it FUSE} spectra of \ion{N}{1}, \ion{Si}{2}, \ion{P}{2}, \ion{Ar}{1}, and \ion{Fe}{2} --- which trace the predominantly neutral (\ion{H}{1}) Galactic and SMC gas. 
We used the component structure determined from moderately high resolution optical spectra of \ion{Na}{1} to model the numerous lines from Galactic H$_2$.
The profile fitting analysis yielded total column densities for those various species in both the Galactic and SMC gas, as well as estimates for the resolution and wavelength accuracy characterizing the {\it FUSE} spectra.

Both the Galactic and SMC gas seem to be predominantly neutral, though a significant fraction of the SMC gas is ionized.
Argon and nitrogen may be partially ionized in the lower column density clouds, particularly in the SMC.
Within the uncertainties in the profile analysis, the overall column densities of \ion{P}{2}, \ion{S}{2}, and \ion{Ar}{1} are consistent with essentially solar ratios, relative to $N$(\ion{Zn}{2}), in both the Galactic and SMC gas.
Accurate abundances for those three elements have not previously been available for the SMC ISM.
\ion{N}{1} may be underabundant, in both Galactic and SMC gas, but observations of weaker \ion{N}{1} lines below 980 \AA~ are needed to refine the abundances and uncertainties as to the total N abundance in the SMC need to be resolved.
Higher resolution spectra of \ion{N}{1}, \ion{O}{1}, and \ion{S}{2} would enable more accurate abundances to be determined for these (and other) mildly depleted species, and would provide better constraints on the ionization of N in the lower column density components.
Using the previously determined \ion{Fe}{2} component structure and column densities, we have estimated $f$-values for nine far-UV \ion{Fe}{2} lines.
For five of the lines, our values agree with those found by Howk et al. (2000); for the other four, our values are 30--40\% lower.

Molecular hydrogen is clearly present in the Galactic gas, with a fractional abundance log[f(H$_2$)] $\sim$ $-$4.3$\pm$0.2, a rotational temperature $T_{01}$ $\sim$ 100 K, and a somewhat higher excitation temperature $T_{\rm ex}$ for the higher $J$ levels --- all similar to values found in low mean density Galactic lines of sight and in the Galactic gas toward several other LMC and SMC stars.
We also report a tentative detection of H$_2$ (from $J$ = 1 and $J$ = 3 only) in the SMC gas, with log[f(H$_2$)] $\sim$ $-$5.7 and $T_{\rm ex}$ of order 1000 K for all rotational levels --- similar to the H$_2$ detected toward $\zeta$ Pup.

Strong absorption from \ion{N}{3}, \ion{S}{3}, and \ion{Fe}{3} has revealed a significant ionized component, particularly in the SMC.
While it is not yet clear how the neutral and ionized gas in the SMC are related, the line profiles suggest that the lower column density, higher velocity SMC components may have a higher fraction of ionized gas.
Strong absorption from \ion{C}{4} and \ion{Si}{4} and (relatively) weaker absorption from \ion{O}{6} are also present at both Galactic and SMC velocities.
The Galactic \ion{O}{6} toward Sk 108 is weaker than that found for any of the halo sightlines reported by Savage et al. (2000).
The $N$(\ion{C}{4})/$N$(\ion{O}{6}) ratio varies somewhat within the SMC --- suggesting that several processes may contribute to the observed high ion abundances.

\section{Acknowledgements}

This work is based on early release/commissioning phase data obtained for the Guaranteed Time Team by the NASA-CNES-CSA {\em FUSE} mission operated by the Johns Hopkins University.  
We are grateful to the entire {\em FUSE} operations team for all their efforts in taking the spectra of Sk 108 and to the data analysis team for their hard work and help throughout the many pipeline iterations required for the analysis of the spectra.  
We thank Chris Howk for useful comments regarding \ion{O}{6}.
Financial support to U.S. participants has been provided by NASA contract NAS5-32985 with Johns Hopkins University.  
Work at the University of Chicago and other institutions listed is funded by subcontracts from JHU.
DEW acknowledges support from NASA LTSA grant NAG5-3228.

\newpage

\begin{figure}
\epsscale{0.8}
\plotone{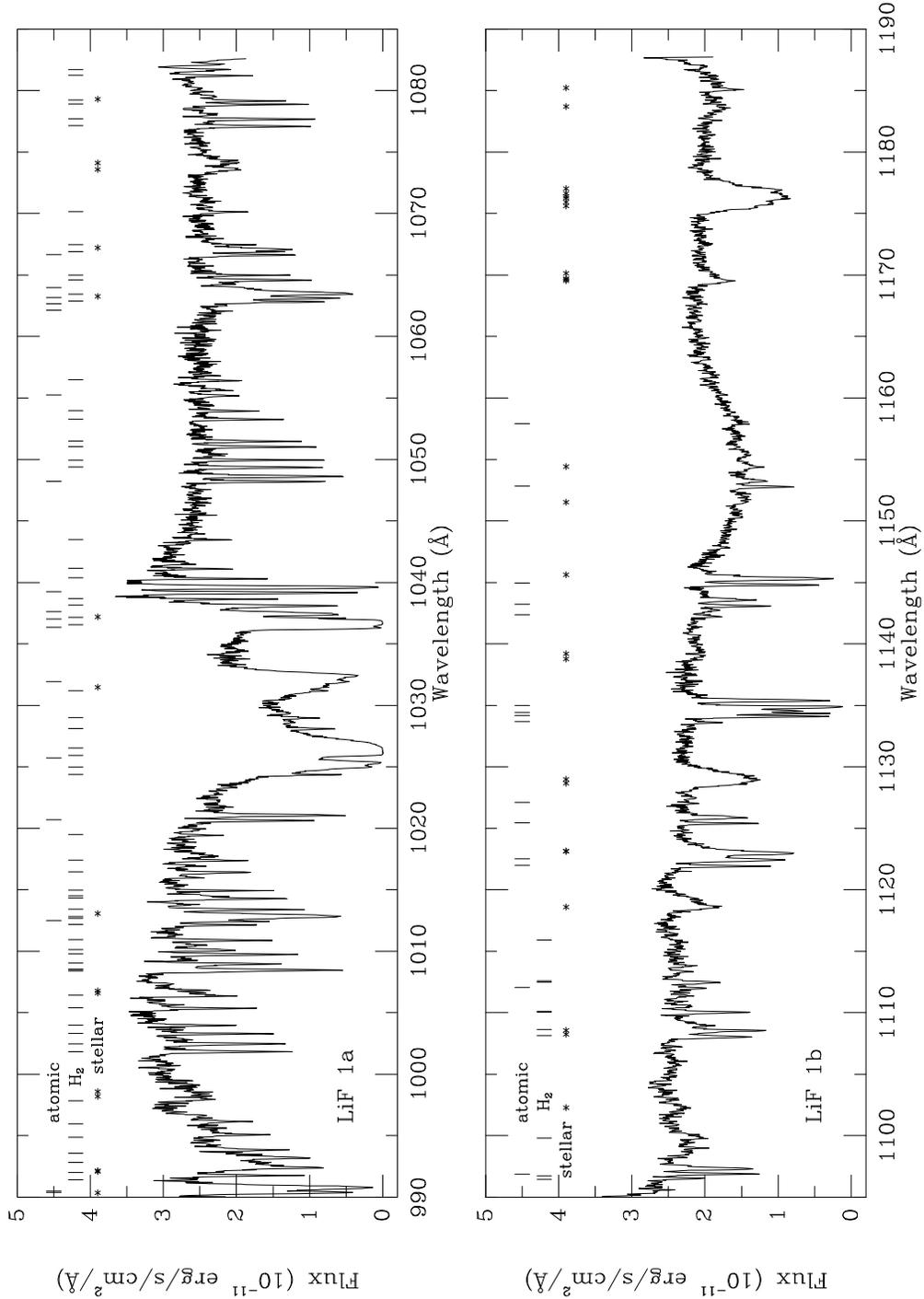}
\caption{
The {\em FUSE} spectrum of Sk 108, binned by 5 pixels.
The two sets of tick marks note the narrow Galactic interstellar absorption lines (H$_2$ and all others); any SMC absorption is typically about 0.5 \AA~ to the red.
The asterisks mark the broader stellar absorption features (Morton \& Underhill 1977); the very broad, shallow feature centered near 1155 \AA~ is an optical anomaly (see text). }
\label{fig:fullspectrum}
\end{figure}

\begin{figure}
\epsscale{0.9}
\plotone{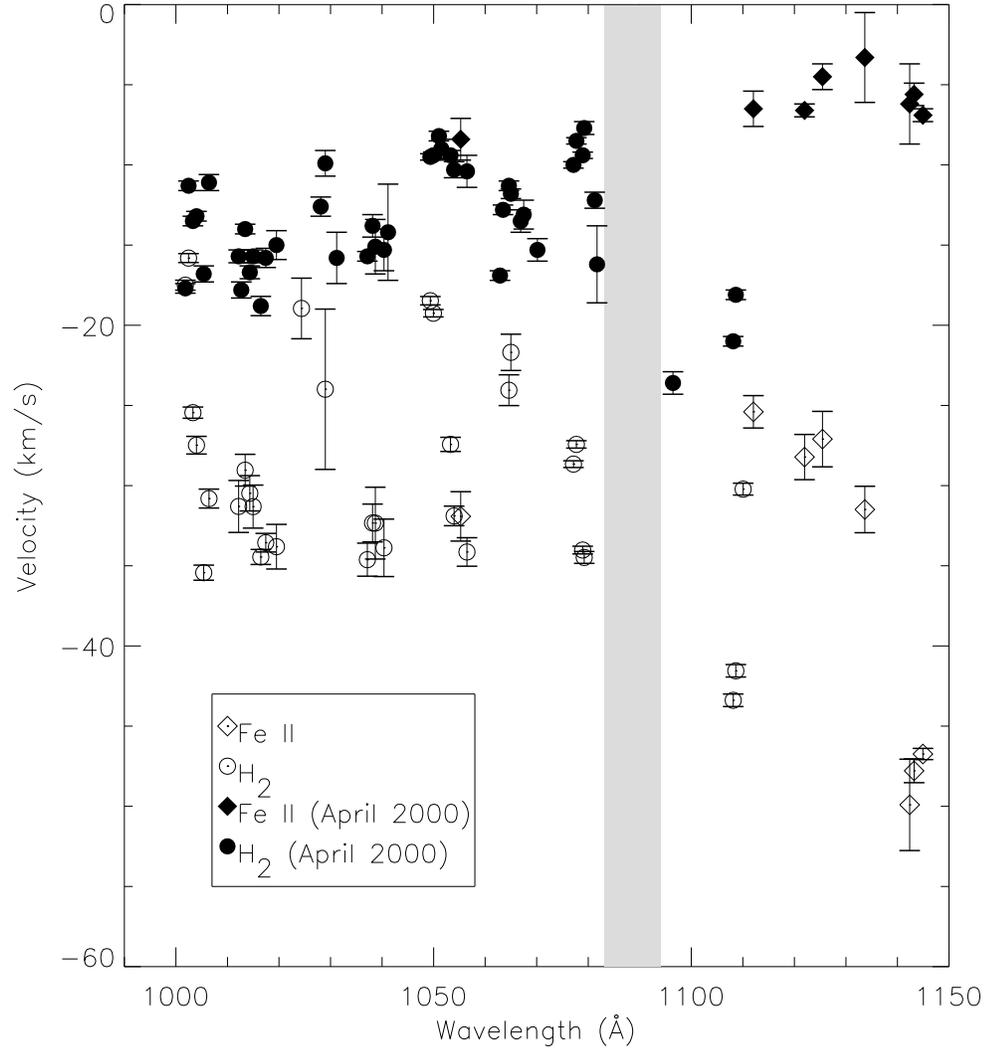}
\caption{
Velocity residuals after profile fitting for Fe II (open diamonds) and H$_{2}$ lines (open circles).  
The grey areas signify the regions in the wavelength space not covered by LiF1A and LiF1B.  
The error bars are approximately 1$\sigma$ uncertainties determined from the profile fits.  
Solid symbols give residuals for an improved wavelength solution (2000 April). 
}
\label{fig:wave}
\end{figure}

\begin{figure}
\epsscale{0.8}
\plotone{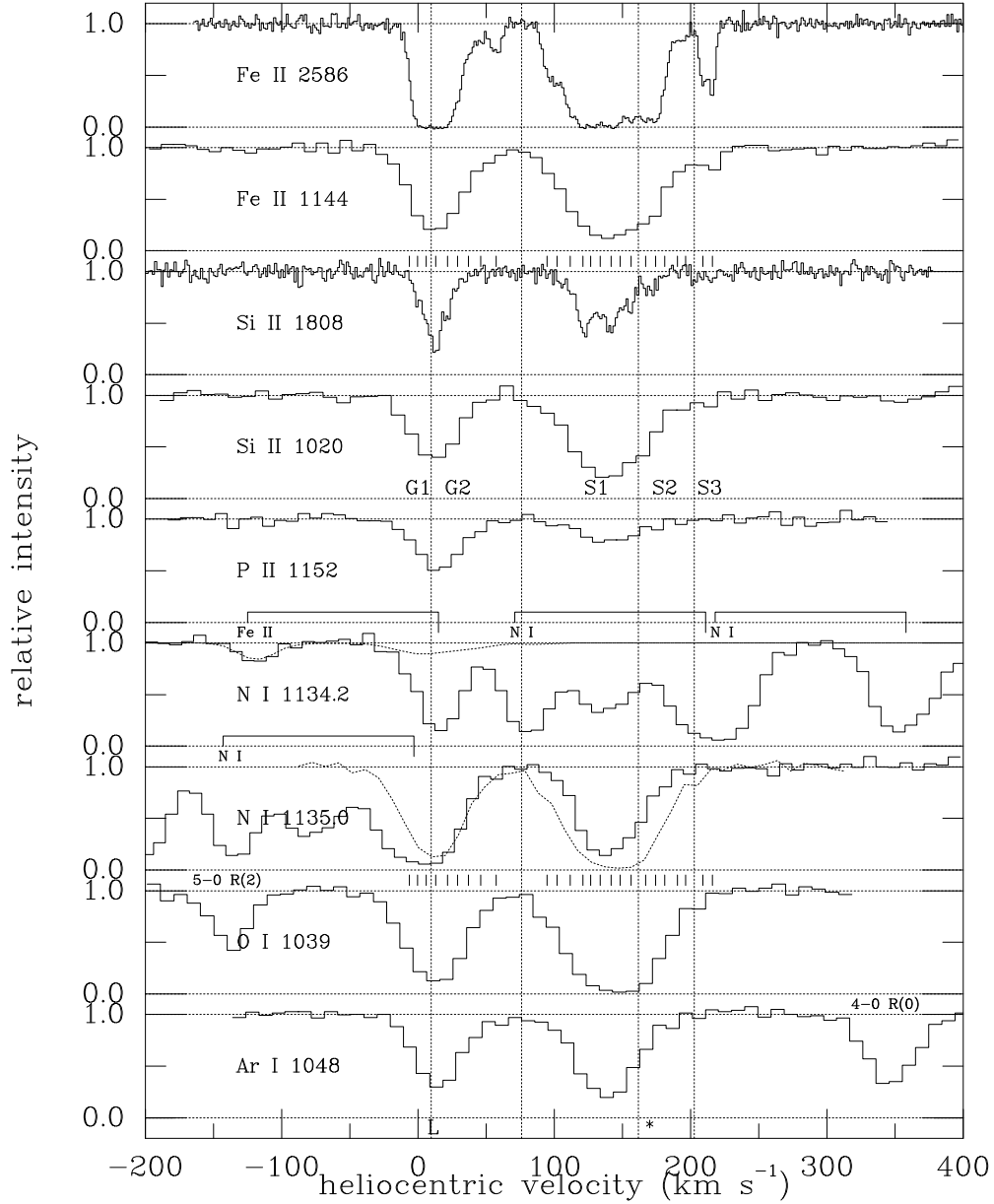}
\caption{
Absorption line profiles for selected neutral and singly ionized atomic species toward Sk 108.  
Profiles for Fe II $\lambda$2586 and Si II $\lambda$1808 are GHRS echelle-B data (R $\sim$ 70,000) (Welty et al. 1997).
All others were obtained with {\em FUSE}, at a resolution of about 12,000.  
The 25 components discerned in the GHRS spectra are shown above the Si II $\lambda$1808 and O I $\lambda$1039 profiles; the G1 and G2 components are Galactic; the S1, S2, and S3 components are SMC.  
The dotted line in the N I $\lambda$1135.0 profile is the O I $\lambda$1039 profile; the dotted line in the N I $\lambda$1134.2 profile shows the predicted absorption from Fe II $\lambda$1133.
The ``L'' at 11 km s$^{-1}$ denotes $v_{LSR}$ = 0 km s$^{-1}$; the asterisk at 170 km s$^{-1}$ denotes the stellar velocity.
}
\label{fig:neutral}
\end{figure}

\begin{figure}
\epsscale{0.8}
\plotone{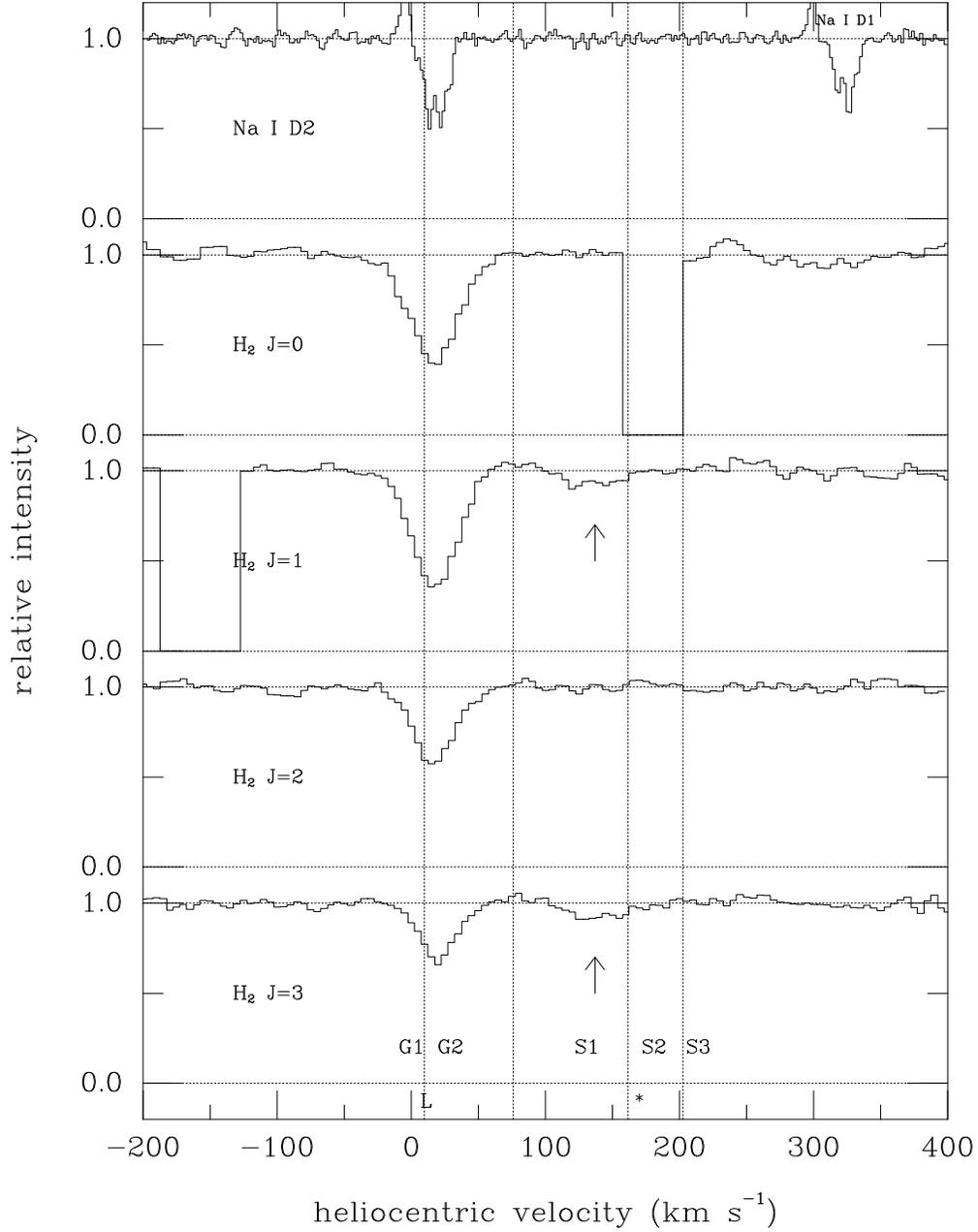}
\caption{
Absorption line profiles for Na I D2 ($\lambda$5889) and several H$_2$ rotational levels toward Sk 108.  
The Na I spectrum was obtained with the AAT UCLES, at a resolution of 5 km s$^{-1}$.  
The H$_2$ profiles for each $J$ are simple averages of three individual {\it FUSE} profiles of similar strength [$J$=0: 2-0 R(0), 4-0 R(0), 8-0 R(0); $J$=1: 2-0 R(1), 4-0 R(1), 8-0 R(1); $J$=2: 2-0 R(2), 4-0 R(2), 8-0 R(2); $J$=3: 4-0 R(3), 5-0 R(3), 6-0 R(3)].  
The relative intensity has been set to zero where absorption from other lines is present in all three of the profiles contributing to the average.  
The G1 and G2 components are Galactic; the S1, S2, and S3 components are SMC.
Note the possible weak absorption at SMC (S1) velocities for $J$=1 and $J$=3 (arrows).
}
\label{fig:h2}
\end{figure}

\begin{figure}
\epsscale{0.8}
\plotone{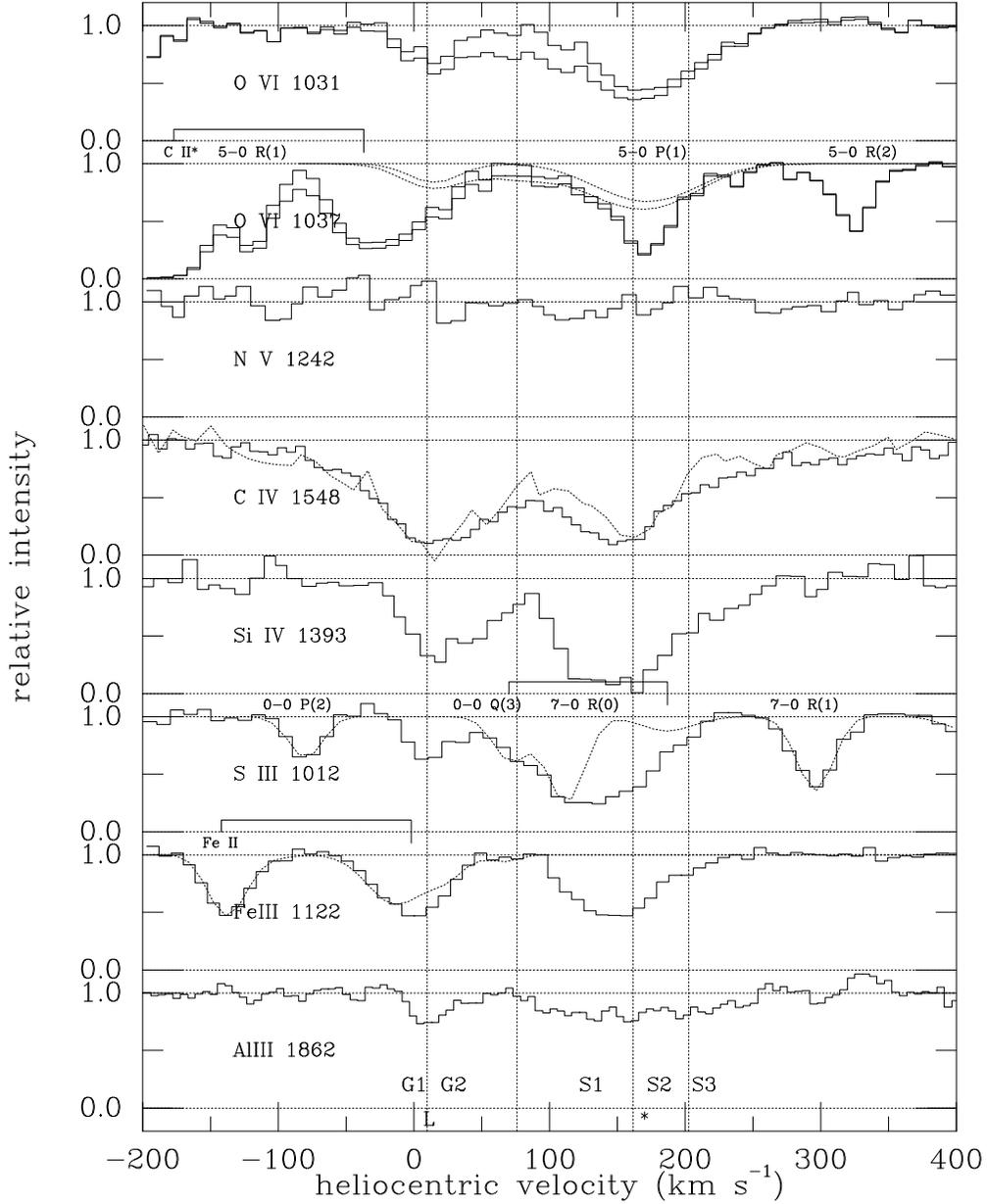}
\caption{
Absorption line profiles for selected lines from higher ions toward Sk 108.  
Profiles for Al III, Si IV, C IV (dotted), and N V are from {\it IUE} (FWHM $\sim$ 20--25 km s$^{-1}$); C IV (solid) is from GHRS (G160M; FWHM $\sim$ 19 km s$^{-1}$; pre-COSTAR, with broad wings in the instrumental profile); O VI, S III, and Fe III are from {\it FUSE} (FWHM $\sim$ 25 km s$^{-1}$).  
Additional species present are noted at velocities corresponding to the Galactic components; SMC C II$^{*}$, H$_2$ 0--0 Q(3), and Fe II are noted (approximately) by longer right-hand tick marks.  
The two profiles shown for the O VI lines correspond to different choices for the continuum (see text and Fig.~6).
The dotted lines in the O VI $\lambda$1037 profiles show the Galactic and SMC O VI absorption predicted from the O VI $\lambda$1031 line.  
The dotted line in the S III $\lambda$1012 profile shows the Galactic and SMC H$_2$ absorption (four lines).
The dotted line in the Fe III $\lambda$1122 profile shows the Galactic and SMC absorption from the Fe II line at 1121.975 {\mbox \AA}.  
}
\label{fig:ionized}
\end{figure}

\begin{figure}
\epsscale{0.8}
\plotone{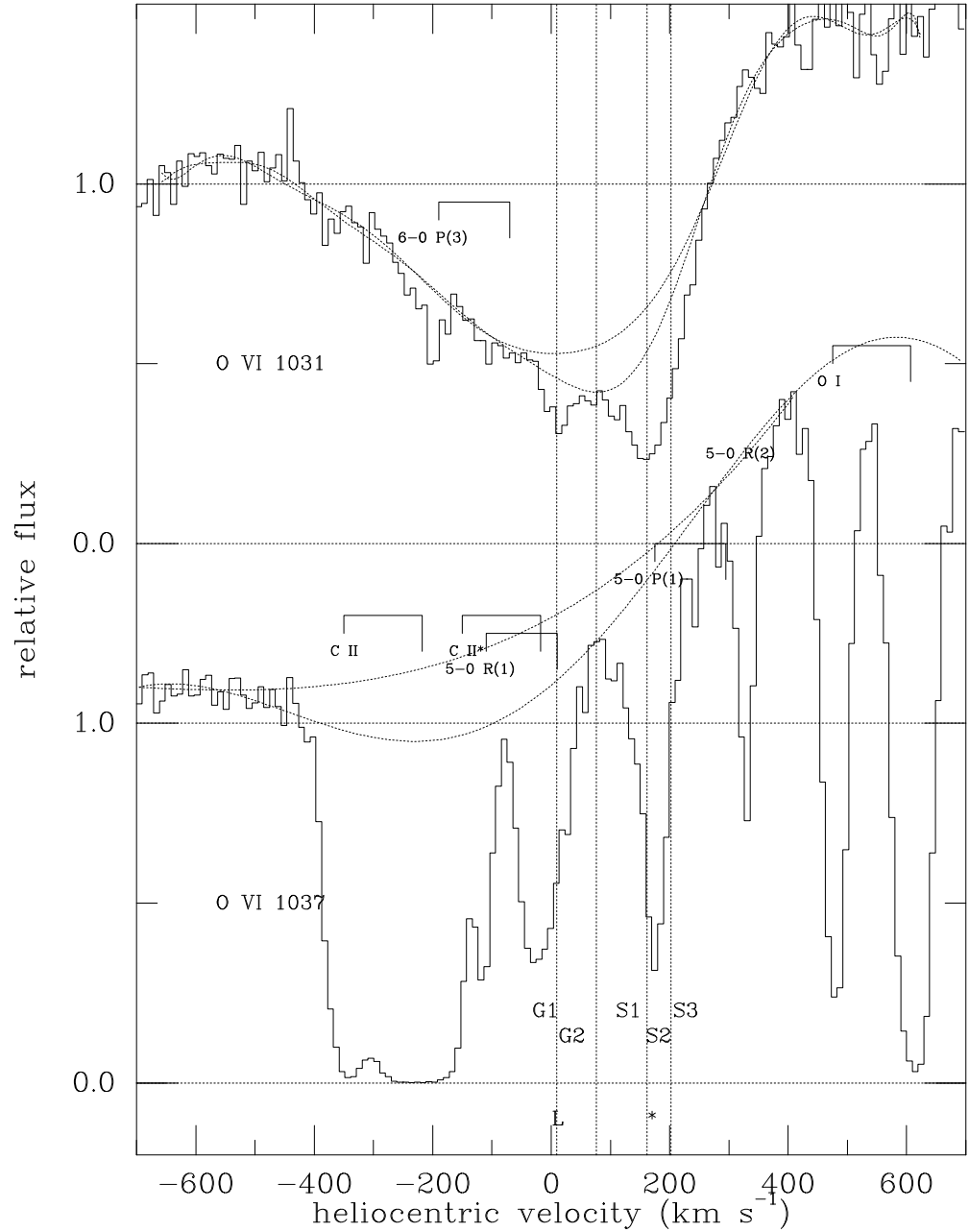}
\caption{
Possible continuum fits for the O VI $\lambda$1031 and $\lambda$1037 regions.  
For O VI $\lambda$1037, the Galactic absorption is blended with SMC C II* and the SMC absorption is blended with Galactic H$_{2}$ 5-0 P(1) (see Fig.~5, second panel from top).} 
\label{fig:o6}
\end{figure}

\newpage

\begin{deluxetable}{lccccccc}
\tabletypesize{\footnotesize}
\tablecolumns{8}
\tablenum{1}

\tablecaption{Equivalent Widths and Column Densities for Neutral and Singly Ionized Species\tablenotemark{a} \label{tab:neutral}}

\tablehead{\colhead{Species} & 
 \colhead{Line\tablenotemark{b}} & \colhead{log($f\lambda$)\tablenotemark{b}} & 
 \multicolumn{2}{c}{Galactic} & \multicolumn{2}{c}{SMC} & 
 \colhead{log($f\lambda$)$_{\rm rev}$\tablenotemark{c}}\\
\cline{4-5} \cline{6-7} \\
\colhead{} &  \colhead{({\mbox \AA})} & \colhead{} & W$_\lambda$\tablenotemark{d} & 
 $N$(X) & W$_\lambda$\tablenotemark{d} & $N$(X) &  \\
\colhead{} & \colhead{} & \colhead{} & 
 \colhead{(m\AA)} & \colhead{(cm$^{-2}$)} & \colhead{(m\AA)} & 
 \colhead{(cm$^{-2}$)} \\
 }

\startdata

N I & 1134.1 & 1.237 & 121$\pm$9 & 15.59$\pm$0.08\tablenotemark{e} & blend & (15.12)\tablenotemark{e} & \nodata \\ 
 & 1134.4 & 1.526 & blend & (15.63)\tablenotemark{e} & blend & (15.11)\tablenotemark{e} & \nodata \\ 
 & 1134.9 & 1.691 & blend & (15.05)\tablenotemark{e} & 175$\pm$12 & 14.90$\pm$0.03\tablenotemark{e} & \nodata \\ 
\\ 
Si II & 1020.6 & 1.461 & 86$\pm$10 & 14.84$\pm$0.05 & 187$\pm$14 & 15.14$\pm$0.04 & \nodata \\ 
& 1808.0 & 0.575 & 94$\pm$11 & 15.37$\pm$0.02 & 158$\pm$20 & 15.52$\pm$0.03 & \nodata \\ 
\\ 
P II & 1152.8 & 2.435 & 76$\pm$11 & 13.71$\pm$0.07 & 47$\pm$15 & 13.25$\pm$0.08 & \nodata \\ 
& 1301.8 & 1.351 & 22$\pm$7  & 13.97$\pm$0.11 & blend & \nodata & \nodata  \\ 
\\ 
S II & 1250.5 & 0.834 & 90$\pm$8 & 15.40$\pm$0.04 & 111$\pm$10 & 15.29$\pm$0.02 & \nodata \\ 
 & 1253.8 & 1.135 & 108$\pm$8 & 15.27$\pm$0.04 & 168$\pm$10 & 15.25$\pm$0.03 & \nodata \\ 
 & 1259.5 & 1.311 & 129$\pm$8 & 15.27$\pm$0.04 & blend & (15.21)& \nodata \\ 
\\ 
Ar I & 1048.2 & 2.430 & 107$\pm$10 & 14.19$\pm$0.06 & 153$\pm$12 & 14.13$\pm$0.04\tablenotemark{e} & \nodata \\ 
 & 1066.6 & 1.834 & 66$\pm$7  & 14.30$\pm$0.05 & blend & (14.25)\tablenotemark{e} & \nodata \\ 
\\ 
Fe II & 1055.3 & 0.926 & 42$\pm$14 & 14.78$\pm$0.11 & 61$\pm$20 & 14.90$\pm$0.09 & 0.96$\pm$0.07 \\ 
 & 1096.9 & 1.545 & 75$\pm$7  & 14.56$\pm$0.04 & 122$\pm$12 & 14.66$\pm$0.03 & 1.35$\pm$0.02 \\ 
 & 1112.0 & 1.006 & 24$\pm$10 & 14.41$\pm$0.12 &  blend    & \nodata & 0.65$\pm$0.12 \\ 
 & 1122.0 & 1.351 & 76$\pm$7  & 14.75$\pm$0.04 &  blend    & \nodata & 1.32$\pm$0.04 \\ 
 & 1125.4 & 1.093 & 64$\pm$7  & 14.90$\pm$0.04 & 97$\pm$12 & 14.97$\pm$0.04 & 1.21$\pm$0.03 \\ 
 & 1133.7 & 0.833 & 20$\pm$11 & 14.54$\pm$0.13 &  blend    & \nodata & 0.60$\pm$0.13 \\ 
 & 1142.4 & 0.451 & 27$\pm$8  & 15.01$\pm$0.08 & 37$\pm$11 & 15.15$\pm$0.08 & 0.72$\pm$0.06 \\ 
 & 1143.2 & 1.182 & 78$\pm$8  & 14.90$\pm$0.04 & 98$\pm$15 & 14.89$\pm$0.05 & 1.26$\pm$0.03 \\ 
 & 1144.9 & 2.080 &144$\pm$10 & 14.67$\pm$0.06 & 272$\pm$14 & 14.68$\pm$0.03 & 1.94$\pm$0.03 \\ 
 & 2260.8 & 0.742 & 58$\pm$12 &14.78$\pm$0.03& 70$\pm$21 &14.84$\pm$0.03& \nodata \\ 
 & 2586.7 & 2.223 & 381$\pm$13 & 14.78$\pm$0.03 & 779$\pm$15 & 14.84$\pm$0.03 & \nodata \\ 
\\ 
Zn II & 2062.7 & 2.717 & 62$\pm$10 & 12.91$\pm$0.03 & 38$\pm$18 & 12.50$\pm$0.08 & \nodata \\  
 
\enddata
\tablenotetext{a}{Column densities are logarithmic values derived from profile fits; lines with $\lambda$ $>$ 1200 \AA~ from GHRS (Paper I), otherwise from {\it FUSE}; uncertainties in W$_{\lambda}$ are 2$\sigma$; uncertainties in $N$ (based on uncertainties in W$_{\lambda}$) are 1$\sigma$.}
\tablenotetext{b}{Wavelengths and $f$-values are from Morton 1991, with updates from Welty et al. 1999b; Ar I from Federman et al. 1992.}
\tablenotetext{c}{Revised values for log($f \lambda$) for Fe II (see text).}
\tablenotetext{d}{Unidentified fixed pattern noise within a line could cause errors quoted to be underestimated.}
\tablenotetext{e}{Fit using modified component structure (see text).}

\end{deluxetable}

\begin{deluxetable}{lcccccc}
\tabletypesize{\small}
\tablecolumns{7}
\tablenum{2}

\tablecaption{Column Densities and Relative Abundances for Neutral and Singly Ionized Species\tablenotemark{a} \label{tab:abund}}

\tablehead{\colhead{Species} & \colhead{A$_{\sun}$\tablenotemark{b}} & 
 \colhead{A$_{\rm SMC}$\tablenotemark{c}} &  \multicolumn{2}{c}{Galactic} & \multicolumn{2}{c}{SMC} \\
\cline{4-5} \cline{6-7} \\
\colhead{} & \colhead{} & \colhead{} & 
 $N$(X) & [X/Zn] & $N$(X) & [X/Zn] \\
\colhead{} & \colhead{} & \colhead{} &
 \colhead{(cm$^{-2}$)} & \colhead{} & \colhead{(cm$^{-2}$)} \\
 }

\startdata

N I & 7.97 & 6.63/7.3\tablenotemark{d} & $>$15.59 & $>-$0.64 & $>$15.12 & $>-$0.70 \\ 
\\ 
Si II & 7.55 & 7.00 & 15.37$\pm$0.02 & $-$0.44 & 15.52$\pm$0.03 & +0.12 \\ 
\\ 
P II & 5.57 & \nodata & 13.86$\pm$0.07 & +0.03 & 13.40$\pm$0.08 & $-$0.02 \\ 
\\ 
S II & 7.27 & 6.59 & $\ga$15.40 & $\ga$$-$0.13 & $\ga$15.29 & $\ga$+0.03 \\ 
\\ 
Ar I & 6.52 & 5.81 & $\ga$14.30 & $\ga$$-$0.53 & $\ga$14.25 & $\ga$$-$0.17 \\ 
\\ 
Fe II & 7.51 & 6.82 & 14.78$\pm$0.03 & $-$0.99 & 14.84$\pm$0.03 & $-$0.52 \\ 
\\ 
Zn II & 4.65 & 4.04 & 12.91$\pm$0.03 & \nodata & 12.50$\pm$0.08 & \nodata \\  
 
\enddata
\tablenotetext{a}{Column densities are logarithmic values; uncertainties are 1$\sigma$.}
\tablenotetext{b}{Solar system meteoritic abundances from Anders \& Grevesse 1989, and from Grevesse \& Noels 1993, except for Ar, which is from Sofia \& Jenkins 1998 (logarithmic, with H = 12.0).}
\tablenotetext{c}{SMC abundances, generally from Russell \& Dopita 1992; see Paper I.}
\tablenotetext{d}{First value is from SMC H II regions and supernova remnants (Russell \& Dopita 1992); second value is typical of SMC stellar abundances (Paper I, and references therein).}

\end{deluxetable}

\begin{deluxetable}{lcccccccc}
\tabletypesize{\small}
\tablecolumns{9}
\tablenum{3}

\tablecaption{Column Densities\tablenotemark{a} of Interstellar H$_{2}$ toward Sk 108 and Selected Galactic Stars \label{tab:h2}}

\tablehead{\colhead{} & \multicolumn{2}{c}{Sk 108 (Galactic)\tablenotemark{b}} & \colhead{Sk 108 (SMC)\tablenotemark{b}} & \multicolumn{2}{c}{HD~28497\tablenotemark{c}} & \colhead{$\zeta$ Pup\tablenotemark{d}} & \colhead{$\mu$ Col\tablenotemark{c}} & \colhead{HD~93521\tablenotemark{e}} \\
\cline{2-3} \cline{4-4} \cline{5-6} \cline{7-7} \cline{8-8} \cline{9-9}\\
\colhead{Transition} & \colhead{100 K} & \colhead{1200 K} & \colhead{} & \colhead{A} & \colhead{B} & \colhead{} & \colhead{} & \colhead{} \\
 }

\startdata
$J=0$ & 34$\pm$17 & $13\pm6$ & $<$0.4 & 0.3 & 0.5 & 0.14 & 11.2 & 229 \\ 
$J=1$ & $58\pm26$ & $26\pm11$ & 0.9$\pm$0.2 & 2.5 & 1.7 & 0.79 & 17.4 & 1040 \\ 
$J=2$ & $18\pm5$ & $10\pm3$ & $<$0.45 &1.3 & \nodata & 0.38 & 2.1 & 149 \\ 
$J=3$ & $4.8\pm1.4$ & $3.6\pm1.0$ & 1.0$\pm$0.2 & 3.4 & 0.8 & 0.92 & 1.0 & 83 \\ 
$J=4$ & $<0.5$ & $<0.5$ &$<$0.5 & 0.7 & $<$0.1 & 0.23 & $<0.1$ & $<2$ \\   
$J=5$ & $<0.6$ & $<0.6$ &$<$0.6 & 0.8 & 0.2 & 0.35 & $<$0.1 & \nodata \\ 
\\ 
log[$N$(H$_2$)] & 16.1 & 15.7 & 14.5 & 15.3 & \nodata & 14.5 & 15.5 & 17.2 \\ 
log[$N$(H I)]\tablenotemark{f} & 20.5 & 20.5 & 20.5 & 20.2 & \nodata & 20.0 & 19.9 & 20.1 \\ 
log[f(H$_2$)] & $-$4.1 & $-$4.5 & $-$5.7 & $-$4.6 & \nodata & $-$5.3 & $-$4.1 & $-$2.6 \\ 

\enddata
%\footnotesize
\tablenotetext{a}{Column densities are all times 10$^{14}$ cm$^{-2}$.}
\tablenotetext{b}{Values for Galactic absorption are from fits with $T$ $=$ 100 K ($b$ $\sim$ 1.5 km s$^{-1}$) or $T$ $=$ 1200 K ($b$ $\sim$ 3.3 km s$^{-1}$) for all 5 components seen in \ion{Na}{1}; uncertainties are rms deviations for the various transitions fitted.
Values for SMC absorption are from simultaneous one component fits to several transitions; uncertainties are approximately 1$\sigma$.}
\tablenotetext{c}{HD 28497 and $\mu$ Col data from Shull \& York 1977.}
\tablenotetext{d}{$\zeta$ Pup data from Morton 1978.}
\tablenotetext{e}{HD 93521 data from Caldwell 1979.}
\tablenotetext{f}{Sk 108 values from Fitzpatrick 1985; all others from Bohlin et al. 1978.}

\end{deluxetable}

\begin{deluxetable}{lccccccc}
\tabletypesize{\small}
\tablecolumns{8}
\tablenum{4}

\tablecaption{Equivalent Widths and Column Densities for Higher Ions \label{tab:ionized}}

\tablehead{\colhead{} & \colhead{} & \colhead{} & \colhead{} &
 \multicolumn{2}{c}{Galactic} & \multicolumn{2}{c}{SMC} \\
\cline{5-6} \cline{7-8} \\
\colhead{Ion} & \colhead{$\lambda$\tablenotemark{a}} & 
 \colhead{log($f \lambda$)\tablenotemark{a}} & \colhead{Source} &
 \colhead{W$_\lambda$\tablenotemark{b}} & \colhead{log[$N$(X)]\tablenotemark{c}} &
 \colhead{W$_\lambda$\tablenotemark{b}} & \colhead{log[$N$(X)]\tablenotemark{c}} \\
\colhead{} & \colhead{(\AA)} & \colhead{} & \colhead{} & \colhead{(m\AA)} & 
 \colhead{(cm$^{-2}$)} & \colhead{(m\AA)} & \colhead{(cm$^{-2}$)}
}

\startdata
Al III & 1862.8 & 2.716 & {\it IUE} & 51$\pm$18 & 12.8 & 150$\pm$40 & 13.3 \\ 
Fe III\tablenotemark{d} & 1122.5 & 1.947 & {\it FUSE} & (33$\pm$7) & 13.8 & 157$\pm$12 & 14.5 \\ 
S III\tablenotemark{d}  & 1012.5 & 1.556 & {\it FUSE} &  52$\pm$13 & 14.3 &(148$\pm$16)& 14.8 \\ 
C IV   & 1548.2 & 2.470 & {\it HST} & 544$\pm$31\tablenotemark{e} & 14.4 & 597$\pm$39\tablenotemark{e} & 14.4 \\ 
       & 1550.8 & 2.169 & {\it HST} & 412$\pm$32\tablenotemark{e} & 14.5 & 463$\pm$38\tablenotemark{e} & 14.5 \\ 
Si IV  & 1393.8 & 2.855 & {\it IUE} & 229$\pm$39 & 13.6 & 466$\pm$49 & 14.1 \\ 
N V    & 1242.8 & 1.986 & {\it IUE} &\nodata & \nodata &   $<$50 & $<$13.7 \\ 
O VI\tablenotemark{d}   & 1031.9 & 2.137 & {\it FUSE} &  51$\pm$18 & 13.7 & 181$\pm$25 & 14.3 \\ 
                        & \nodata&\nodata& {\it FUSE} & 93$\pm$19 & 14.0 & 267$\pm$28 & 14.5 \\ 
\enddata
\tablenotetext{a}{Morton 1991.}
\tablenotetext{b}{Uncertainties are 2$\sigma$.}
\tablenotetext{c}{Column density estimates from apparent optical depth integral --- may be underestimates for strong lines.  Typical errors are $\pm$0.1 dex, varying somewhat for different species.}
\tablenotetext{d}{Stellar lines removed.  Some stellar absorption may be present for Si IV.  Two values for O VI correspond to maximum and minimum stellar features assumed (see Figures 5 and 6).}
\tablenotetext{e}{Hutchings et al. 1993 cite 570 m\AA~ for $\lambda$1548 and 410 m\AA~ for $\lambda$1550 (both Galactic and SMC for each).}

\end{deluxetable}

\end{document}